\documentstyle[jkas]{article}
 
\runningauthor{ANN \& HWANG}
\runningtitle{BULGES OF BARRED GALAXIES}
\beginpage{1}
\endpage{6}

\def\etal{{et al.}\ }

\def\s-1{{\rm\,s^{-1}}}

\def\spose#1{\hbox to 0pt{#1\hss}}
\def\lta{\mathrel{\spose{\lower 3pt\hbox{$\mathchar"218$}}
     \raise 2.0pt\hbox{$\mathchar"13C$}}}
\def\gta{\mathrel{\spose{\lower 3pt\hbox{$\mathchar"218$}}
  \raise 2.0pt\hbox{$\mathchar"13E$}}}

\begin{document}

\font\twelvei = cmmi10 scaled\magstep1
\font\teni = cmmi10 \font\seveni = cmmi7
\font\mbf = cmmib10 scaled\magstep1
\font\mbfs = cmmib10 \font\mbfss = cmmib10 scaled 833
\font\msybf = cmbsy10 scaled\magstep1
\font\msybfs = cmbsy10 \font\msybfss = cmbsy10 scaled 833
\textfont1 = \twelvei
\scriptfont1 = \twelvei \scriptscriptfont1 = \teni
\def\mit{\fam1 }
\textfont9 = \mbf
\scriptfont9 = \mbfs \scriptscriptfont9 = \mbfss
\def\bmit{\fam9 }
\textfont10 = \msybf
\scriptfont10 = \msybfs \scriptscriptfont10 = \msybfss
\def\bmsy{\fam10 }

\def\eg{{\it e.g.}}
\def\ie{{\it i.e.}}
\def\lsim{\raise0.3ex\hbox{$<$}\kern-0.75em{\lower0.65ex\hbox{$\sim$}}}
\def\gsim{\raise0.3ex\hbox{$>$}\kern-0.75em{\lower0.65ex\hbox{$\sim$}}}

\title{BULGES OF TWO BARRED GALAXIES: NGC~3412 AND NGC~3941}
\author{
{\large\textbf{\textsc{H}}}\textbf{\textsc{ONG}}
{\large\textbf{\textsc{B}}}\textbf{\textsc{AE}}
{\large\textbf{\textsc{A}}}\textbf{\textsc{NN}}
\textbf{\textsc{AND}}
{\large\textbf{\textsc{I}}}\textbf{\textsc{NOK}}
{\large\textbf{\textsc{H}}}\textbf{\textsc{WANG}}
}
   
\address{Department of Earth Sciences, Pusan National University, Pusan
   609-735, Korea\\
{\it E-mail:hbann@cosmos.es.pusan.ac.kr}}
\address{\normalsize{\it (Received Nov. 20, 1998; Accepted Feb. 1, 1999)}}

\received{Nov. 20, 1998}
\accepted{Feb. 1, 1998}

\abstract{
We have conducted near-infrared ($J$- and $H$-band) 
surface photometry for two early type barred
galaxies, NGC~3412 and NGC~3941.
The bulges of NGC~3412 and NGC~3941 show isophotal twists
which indicate that they are triaxial. NGC~3412 has a very short bar and its
bulge is more centrally concentrated than that of NGC~3941.
The unusually short bar and the centrally concentrated triaxial
bulge of NGC~3412 might be the result of bar dissolution.
The colors of the nuclear region of NGC~3941 resemble those
of the blue nuclei, implying the presence of young stellar populations.
}

 
\maketitle

\section{INTRODUCTION}

There have been many observational evidences that bulges may not be simply
described by scale-down ellipticals.  A large number of barred spirals
are known to 
have triaxial bulges (Kormendy 1979; Ann 1995).
The luminosity distribution of
late type bulges are better fitted by exponential function than the 
$r^{1/4}$ law (Courteau, de Jong, \& Broeils 1996).  Recent $HST$ WFPC2 images
show that highly irregular bulges are not so unusual (Carollo \etal 1997).

The morphology of bulges has been analyzed by the shape of
isophotes.  Earlier studies of the isophotes of M31 lead to the conclusion
that the bulge of M31 is triaxial because of the presence of isophotal twists
(Lindblad 1956; Stark 1977).  The 
isophotal twists can be caused by
triaxial structures in galaxies (Bertola, Vietri, \& Zeilinger 1991;
Wozniak \etal 1995).  Recent survey of the isophotal maps
of barred galaxies (Ann 1995) showed that more than half of the 104 bright
barred galaxies have triaxial bulges.  
Elmegreen \etal (1996) found that 
64 \% of 80 SB and SAB galaxies show isophotal twists from the survey of 
galaxies at near-infrared.

There are several mechanisms which can account for the isophotal twists
observed in the inner regions of barred galaxies.  Apart from the triaxial
bulges mentioned above, the presence of two inner Lindblad resonances (ILRs),
which affect orbits of stars and gas clouds (Shaw \etal 1993), and the
secondary bars (Friedli \& Martinet 1993) may distort the isophotes. 
Recent N-body simulations by Norman, Sellwood, \& Hasan (1996) showed
that the triaxial bulges are remnants of bar dissolution. 

NGC~3412 and NGC~3941 are early type barred galaxies of which bulges are
found to be triaxial due to the misalignments between the major axes
of bulge and bar (Shaw \etal 1995).  However, because of small 
misalignment between the major axes of bulge and disk, the bulge of NGC~3941
was classified as an oblate spheroid by Ann (1995). The colors of the 
nucleus of NGC~3412 are known to be bluer than those of the
surrounding regions (Shaw \etal 1995).
The purpose of the present study is to analyze the bulge morphology by
near-infrared images and to draw some insights on the origin of the triaxial
bulges.  The basic parameters of NGC~3412 and NGC~3941 are given in Table~1.

\begin{table*}
\begin{center}
{\bf Table 1.}~~Basic data of NGC~3412 and NGC~3941
\vskip 0.3cm
\begin{tabular}[t]{ccc}
\hline\hline
& NGC 3412 & NGC 3941\\
\hline
 R.A.(1950)                 &  $10^{h}48^{m}14.^{s}50$  &  $11^{\circ}50'19."67$ \nl
 Dec(1950)                  &  $13^{h}40^{m}41.^{s}00$  &  $37^{\circ}15'52."00$ \nl
 Morphological type         &  SBO                      &  SBO                   \nl
 logD$_{25}$                &  1.56                     &  1.54                  \nl
 logR$_{25}$                &  0.25                     &  0.18                  \nl
 Total magnitude(B$_T$)     &  11.45                    &  11.25                 \nl
 Galactic Extinction(B mag) &  0.06                     &  0.00                  \nl
 (B-V)$_T$                  &  0.91                     &  0.91                  \nl
 (U-B)$_T$                  &  0.39                     &  0.44                  \nl
 $V_H$                      &  $865 \pm 27 km/s $       &  $928 \pm 5 km/s$      \nl
\hline
\end{tabular}
\end{center}
\end{table*}

  In Section II, we present the observations and data reduction.  
The morphology and the luminosity distributions are described 
in Section III, and discussion and brief summary are given in the final section.

\section{OBSERVATIONS AND DATA REDUCTION}

We conducted a $JH$ imaging of NGC~3412 and NGC~3941 with
a PtSb $256 \times 256$ IR array attached to the modified Newtonian
focus of the 1.8 m Plaskett Telescope of DAO in 1993 April.  The pixel
size is 30 $\mu\rm{m}$ which corresponds to 0.$^{''}$67 on the sky.
The readout noise and the dark of the system were $62 e^{-1}$ and
${7 e^{-1} s^{-1} {\rm pixel}^{-1}}$, respectively. 
The flat-field exposures were done on the twilight sky.
The images were obtained with exposure times of 480 s in {\it J}- 
and {\it H}-band.
Because dark noise is not insignificant, we obtained  
dark frames with various exposure times including those for the 
object frame. 
Several standard stars from Elias \etal (1982)
were observed for the transformation of the instrumental magnitudes to the 
standard system.  The seeing during the observation was about 
$3^{\prime \prime}$.

The basic reductions were carried out using the CCDRED package in IRAF.
This procedure includes subtraction of a master bias frame with
overscan correction, dark subtraction, trimming of the data section, 
and flat-fielding.  
Flat-fielding was made by dividing the bias subtracted images by a master 
flat field frame which was prepared by median 
combining of all the flat field frames in each filter. 

The flat-fielded frames of galaxy images were subtracted and divided
by sky frames which were obtained by fitting the sky regions surrounding the
galaxy images, by using IRAF/SPIRAL. 
We applied a variable
width Gaussian beam smoothing to increase the signal-to-noise ratio of the 
luminosity distribution in the outer parts of the galaxy.
The absolute calibration of the galaxy photometry was made by the photometry
of the standard stars.  Because of the large photometric uncertainties,
we derived only the zero points for each filter.

\section{Results}

\subsection{Isophotal Maps}

Fig.~1 shows the isophotal maps of NGC~3412 (a, b) and 
NGC~3941 (c, d) with $J$-band in the left panel and $H$-band in the 
right. Here the tick intervals are 10 pixels ($6.^{\prime \prime}7$) each
and the contour interval is 0.5 mag arcsec$^{-2}$. North is up and east to
the left.
As can be seen in Fig~1, there is no visible difference between the
$J$- and $H$-band isophotes except for the innermost isophotes which are
heavily affected by the atmospheric seeing. The shapes of the 
bulges of NGC~3412 and NGC~3941 are similar to each other.
They are somewhat elongated with their 
position angles of major axes being different from those of the bars. 
There is no clear distinction between the bulge and the bar of NGC~3412
in  their isophotes, but the weak variation of ellipticities and position
angles, as shown in Fig.~2, are due to the presence of bar which ends near
$r \sim 13^{\prime \prime}$.  The bar of NGC~3941 is easily distinguishable
from the bulge and disk in their isophotal maps.

\subsection{Luminosity Distribution}

The surface brightness profiles, along with the profiles of ellipticity 
and position angle, derived from ellipse fittings to the observed
isophotes are plotted in Fig.~2.  We used the ellipse fitting program
in IRAF/SPIRAL which is essentially the same as that described by
Kent (1983).  There seems to be no difference between the surface brightness
profiles of $J$- and $H$-band for NGC~3412 (Fig.~2-(a))
but the $J$-band profile of NGC~3941 is steeper than the $H$-band profile
in the bulge dominated region of $r < 10^{\prime \prime}$ (Fig.~2-(b)).
The steep gradient in $H$-band makes the
colors of the inner region bluer than the outer region.
The bar of NGC~3412 is difficult to be recognized in the surface brightness
profiles due to negligible bar luminosity but the bar of NGC~3941
is easily identified by a clear bump in the surface brightness profiles.

The profiles of ellipticity and position angle are presented in Fig.~2-(c)
and (e) for NGC~3412 and Fig.~2-(d) and (f) for NGC~3941, respectively.
In general there is no big difference in the radial variations of  
ellipticities and position angles in different passbands except for those
at $r > 25^{\prime \prime}$ in NGC~3412 and at $r < 5^{\prime \prime}$ in
NGC~3941. As shown in Fig.~2-(c) and (d) the mean ellipticity of the bulge
of NGC~3412 is $\approx 0.2$ with little variation along the radius while 
the ellipticity of NGC~3941 varies smoothly from 0.2 to 0.4 in the bulge 
with a mean value of $\sim 0.3$.
The reason for the change of the bulge ellipticities in NGC~3941 is due to
the bar luminosity which dominates the observed luminosity 
near $r \sim 20^{\prime \prime}$.

The bulges of NGC~3412 and NGC~3941 have been classified as triaxial 
because of the isophotal twists in the $JHK$ images (Shaw \etal 1995). 
The smooth variation of position angles of NGC~3941 shown in Fig.~2-(f)
indicates that the bulge of NGC~3941 is triaxial.  But if we
consider that the position angle of the bulge of NGC~3941 is similar to that
of the disk, there is a possibility of an oblate spheroid for the bulge of 
NGC~3941.  Because Ann (1995) took into account the misalignments between
three components (bulge, disk, and bar) together, he classified the
bulge of NGC~3941 as an oblate spheroid.  The variation
of position angles of NGC~3412 in Fig.~2-(e) is more complicated.  But the
behavior resembles the third type of isophotal twists defined by
Wozniak \etal (1995), $\it{barred~plus~twisted~isophotes}$, which is 
thought to be caused by bar and triaxial bulge.

\begin{table*}[bt]
\begin{center}
{\bf Table 2.}~~Effective radius and Effective brightness of 
the bulges of NGC~3412 and NGC~3941
\begin{tabular}[t]{cccccccc}
\hline\hline
Galaxy & & {$r_{e}$} & {$\mu_{e}$} &
Galaxy & & {$r_{e}$} & {$\mu_{e}$}  \\
\hline
N3412 &  J  &  13.8  & 18.12  &  N3941  & J &  28.1 &  18.82 \\
      &  H  &  11.5  & 16.95  &         & H &  31.6 &  18.14 \\
\hline
\end{tabular}
\end{center}
\end{table*}

\begin{figure*}
\centerline{\epsfysize=17cm\epsfbox{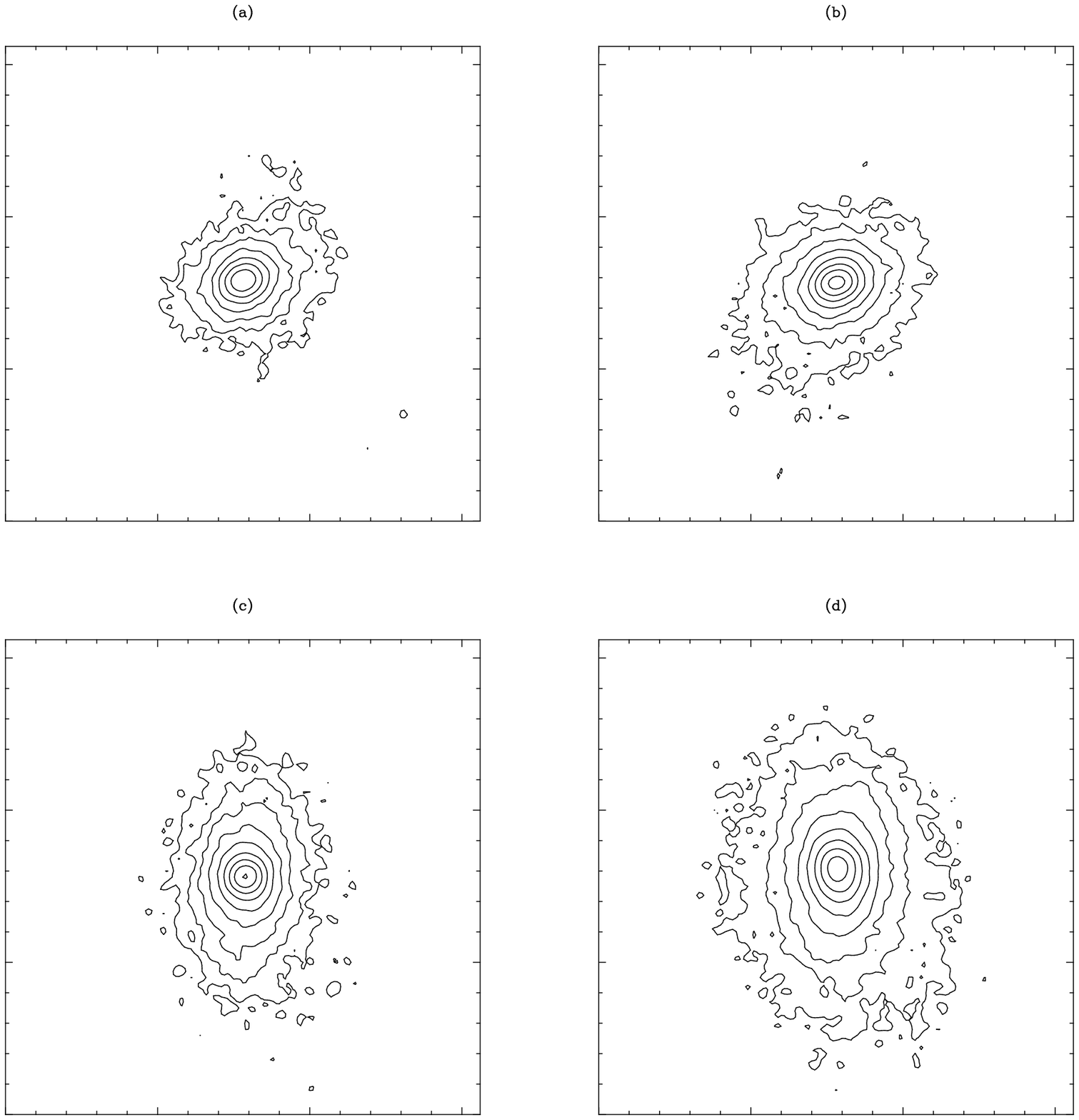}}
\vskip 0.5cm
\hskip 1.1cm {\begin{minipage}{15cm}
{\bf Fig. 1.}---~The $J$- and $H$-band isophotal maps 
of NGC~3412 (a, b)
and NGC~3941 (c, d).  $J$-band images are in the left panels and $H$-band
images are in the right.  The outermost contours 
are 18.5 mag arcsec$^{-2}$ for $J$-band and 18.0 mag arcsec$^{-2}$
for $H$-band, respectively.
\end{minipage} }

\end{figure*}

\begin{figure*}[p]
\centerline{\epsfysize=18cm\epsfbox{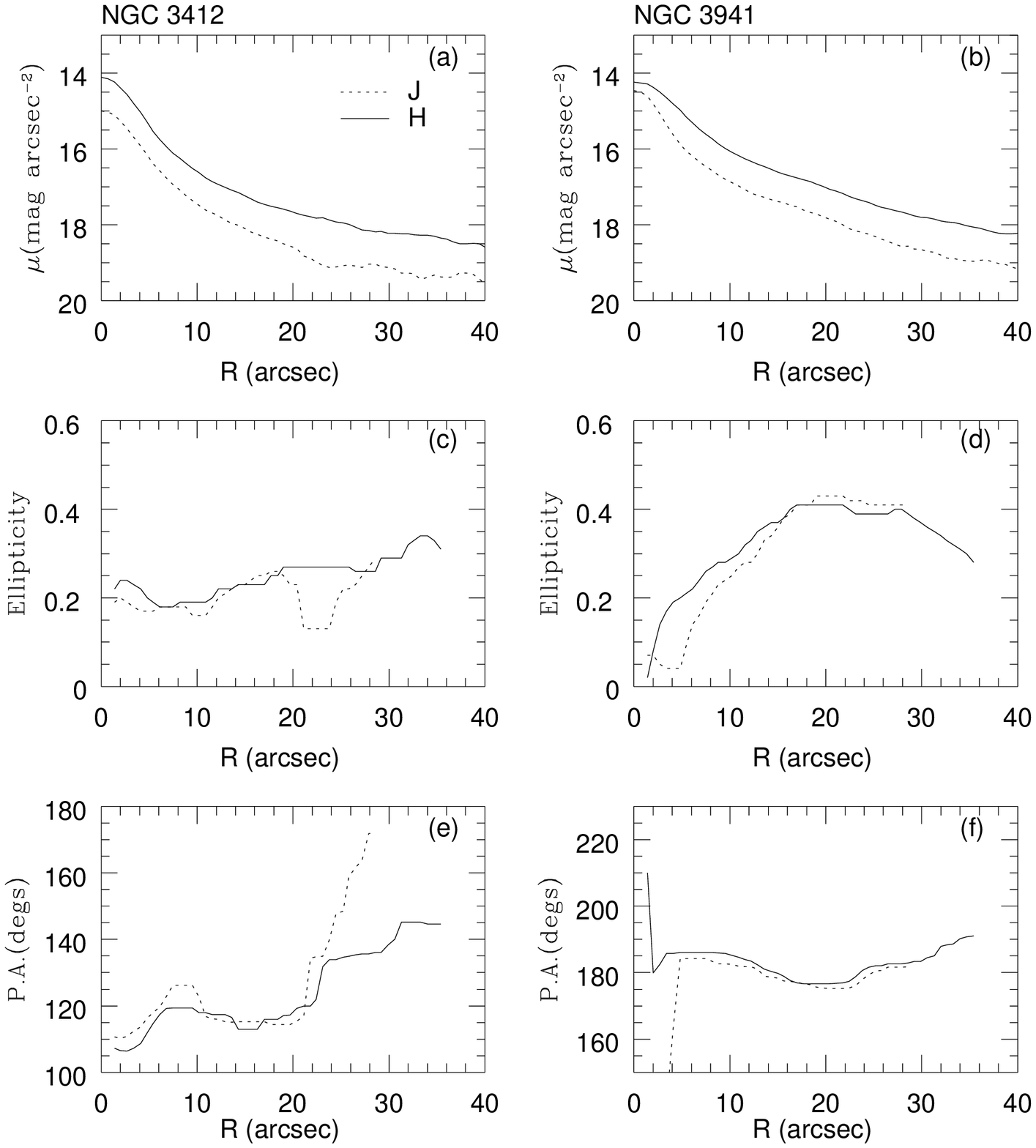}}
\vskip 0.5cm
\hskip 1.1cm {\begin{minipage}{15cm}
{\bf Fig. 2.}---~Profiles of surface brightness, ellipticity and position 
angle (P.A.) from ellipse fittings.  The profiles of NGC~3412 are in the 
left panels and those of NGC~3941 are in the right panels.
\end{minipage}}
\end{figure*}
\newpage

We determined the effective radius ($r_{e}$) and the effective surface 
brightness ($\mu_{e}$) of the bulges of NGC~3412 and NGC~3941 by fitting
the observed profiles to the de Vaucouleurs' $r^{1/4}$-law. We summarize
the derived scale parameters in Table~2. The mean effective radius
of the bulge of NGC~3412 is less than half of NGC~3941.
Because both the isophotal radii and the distances of NGC~3412 and NGC~3941
inferred from the radial velocities in Table~1 are nearly the same,
the bulge of NGC~3412 is intrinsically more centrally concentrated than
that of NGC~3941. The bar luminosity might affect the scale parameters,
especially in NGC~3941. However, the difference of the bulge scale lengths
seems to be real because these parameters were determined where the bulge
luminosities dominate the observed luminosity profiles.
The small differences in the effective radii of the bulges in different
passbands seem to be due to photometric errors.

\subsection{Color Profiles}

Fig.~3 shows the $J-H$ color profiles along the major- and minor axes, together
with the mean color profiles of the concentric
annuli whose radii are defined as $r=\sqrt{r_{1}r_{2}}$ where $r_{1}$
and $r_{2}$ are the radii of the inner and outer annuli, respectively.
General trend of the three color profiles are similar to each other with
some fluctuations due to low signal-to-noise ratio of the photometry.
$J-H$ color of the nuclear region of NGC~3412 is not much different
from that of the surrounding region.  But it is redder than that of the
bar ($\Delta(J-H) \approx 0.1$ mag).  This color distribution is inconsistent 
with the photometry of Shaw \etal (1995) who showed that NGC~3412
has a blue nucleus which is $0.14$ mag bluer than the
surrounding region.  Because of the probable errors in the zero point 
calibrations, the
difference between the nuclear colors of the present photometry and those of
Shaw \etal (1995) is quite plausible, but the opposite trend of the color
variation along the radius is difficult to explain unless the 
photometric errors are very large.  Our $J-H$ color of the bulge of 
NGC~3412 ($\sim 0.7$) is similar to that of the mean $J-H$ color
($0.78 \pm0.03$) of bulges of ordinary spirals (Glass 1984).

The color profile of NGC~3941 is very unusual in the sense that it
shows a very 
blue nucleus with a steep gradient inside $r \sim 8^{\prime \prime}$ with
the largest gradient along the major-axis profile.
This blue nucleus of NGC~3941 was not observed in the $JHK$ photometry of
Shaw \etal (1995) although their photometry shows that 
the nuclear color of NGC~3941 ($J-H=0.53$)
is somewhat bluer than that of the mean color ($0.58 \pm 0.07$) of the 
blue nuclei galaxies.  

\begin{figure*}
\centerline{\epsfysize=10cm\epsfbox{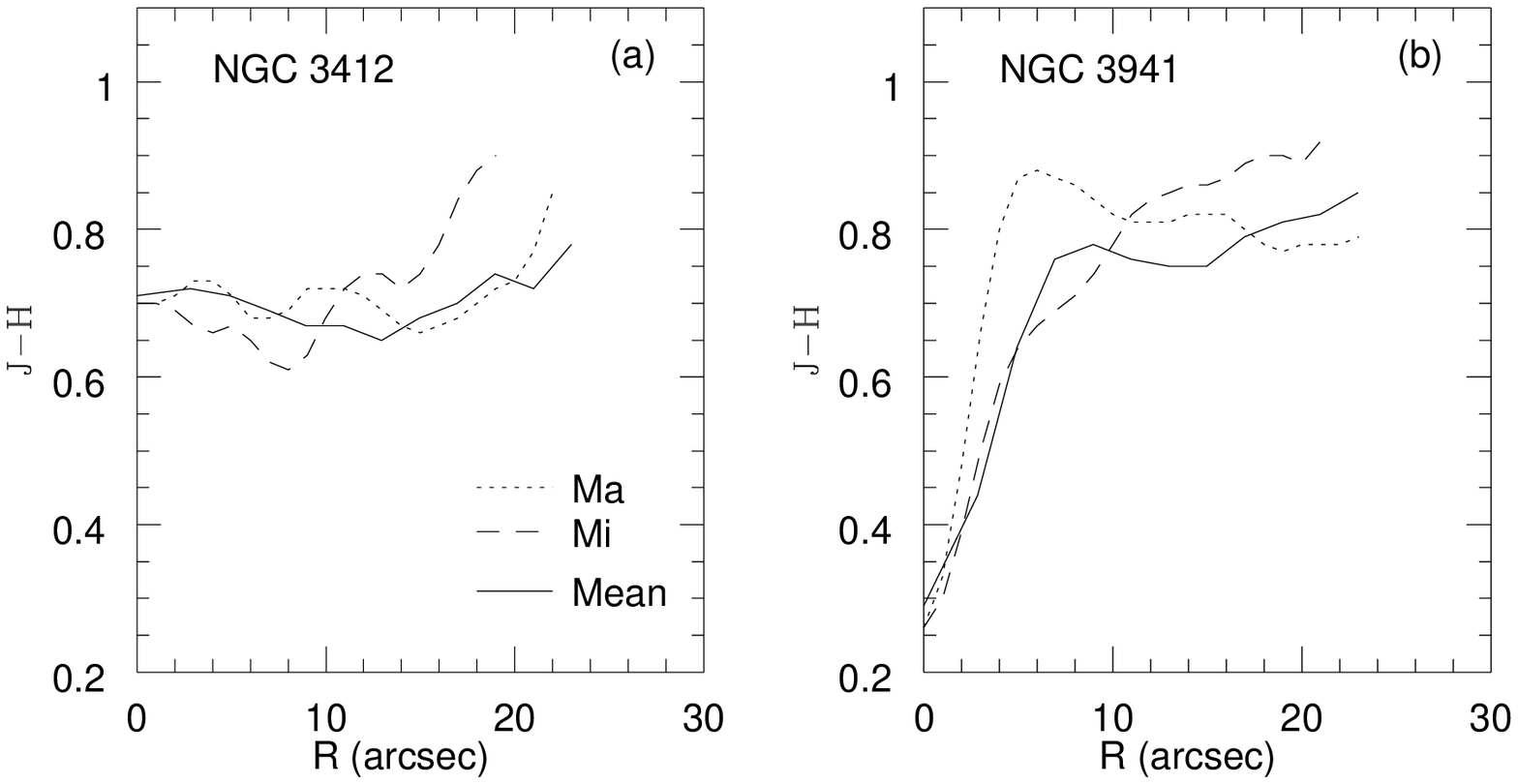}}
\vskip 0.5cm
\centerline{
{\bf Fig. 3.}---~Color profiles of NGC~3412 and NGC~3941.  Notice the blue nucleus 
of NGC~3941.}
\end{figure*}
\vskip 1cm

\section {DISCUSSION AND SUMMARY}

We analyzed the bulge morphology of two barred galaxies NGC~3412 and NGC~3941
by near-infrared photometry.  The present photometry  shows elongated
bulges of both of the galaxies with a mean ellipticity of $\sim 0.2$.
There are misalignments between the major axes of the bulges and bars,
indicating that their bulges are triaxial.
However, as noted by Ann (1995), there is a possibility of an oblate
spheroid for the bulge of NGC~3941 because the position angle of the
bulge is similar to that of the disk. 

The bulge of NGC~3412 is more centrally concentrated than that of NGC~3941.
The small bar and centrally concentrated triaxial bulge of NGC~3412 may 
suggest that the triaxial bulge is the result of bar dissolution.
Recent N-body simulations of Norman,
Sellwood, \& Hasan (1996) showed that centrally concentrated bulges
can destroy bars if the central mass is larger than 5 \% of the 
total mass.  Because our photometry does not extend to the outer parts of
the galaxy, the mass fraction of the bulge of NGC~3412 can not be 
determined by the present photometry.  However, it is quite plausible that 
the mass inside the centrally concentrated bulge is larger than that
required for the bar dissolution if we consider the morphological type of 
NGC~3412 is SB0.

The $J-H$ color of the nuclear region of NGC~3941 is $\sim 0.3$ mag bluer than
those of the surrounding regions.  Blue nuclei similar to the nucleus of
NGC~3941 have been observed for $\sim 70 \%$ of 32 barred galaxies
(Shaw \etal 1985).  Their sample includes NGC~3412 and NGC~3941.  But their 
photometry gives a normal nucleus for NGC~3941 and a blue nucleus 
for NGC~3412. This is opposite to the present result.  We do not
know the reason for the discrepancy between our photometry and that of Shaw 
\etal (1995).  However, if the blue nucleus of NGC~3941 is real, it 
might be caused by the young stellar populations which are formed
by the gas driven by the bar of NGC~3941.  Recent SPH simulations show 
that strong bars are very effective to drive gas inflow which leads to
burst of star formations (Friedli \& Bens 1993; Ann \& Kwon 1996).
If the isophotal twists observed in the bulge of NGC~3941 are caused by the
gas inflow which leads to the burst of star formation that results in the
blue nucleus, the bulge of NGC~3941 is triaxial regardless of the similarity
of the position angles of the bulge and disk.

\acknowledgements
H. B. Ann is grateful to the hospitality provided by DAO during his
the stay at DAO.
This work was supported in part by the Basic Science Research 
Institute Program, Ministry of Education, 1997, BSRI-97-5411.

\newpage
\input epsf

\end{document}